\documentclass[secnumarabic,amssymb,nobibnotes,aps,prd,showpacs]{revtex4}%
\usepackage{graphicx}
\usepackage{dcolumn}
\usepackage{bm}
\usepackage{amsmath}%
\usepackage{amsfonts}%
\usepackage{amssymb}

\begin{document}


\title{Prompt emission from GRB 150915A in the GeV energy range detected at ground by the New-Tupi detector: A review}

\author{C. R. A. Augusto, C. E. Navia, M. N. de Oliveira}
\affiliation{Instituto de F\'{\i}sica, Universidade Federal Fluminense, 24210-346,
Niter\'{o}i, RJ, Brazil} 

\author{Andr\'{e} Nepomuceno}
\affiliation{Departamento de Ci\^{e}ncias da Natureza, IHS, Universidade Federal Fluminense, 28890-000, Rio das Ostras, RJ, Brazil}


\author{V. Kopenkin}
\affiliation{ Research Institute for Science and Engineering, Waseda University, Shinjuku, Tokyo 169, Japan.}

\author{T. Sinzi}
\affiliation{Rikkyo University, Toshima-ku, Tokyo 171, Japan}

\date{\today}
\begin{abstract}

Since 2014, a new detector (New-Tupi) consisting of four plastic scintillators ($150 \times 75 \times 5 cm^3$) placed in pairs and located in Niteroi, Rio de Janeiro, Brazil, has been used for the search of transient solar events and photomuons from gamma-ray bursts (GRBs).
On September 15, 2015, at 21:18:24 UT,  the Swift Burst Alert Telescope (BAT)  triggered and located GRB 150915A (trigger 655721). 
The GRB light curve shows a weak complex structure of long duration  $T_{90}=164.7 \pm 49.7 $ s, and a fluence in the 15-150 keV band of  $8.0 \pm 1.8 \times 10^{-7}erg/cm^2$.  
GRB 150915A was fortuitously located in the field of view of the New-Tupi detector, and a search for prompt emission in the GeV energy range is presented here. 
The analysis was made using the ``scaler'' or ``single-particle'' technique.
The New-Tupi detector registered a counting rate excess peak of duration $T_{90}=(6.1\pm 0.6)$ s with a signal significance $(4.4\pm 0.5)\sigma$, (and not $6.9\sigma$ as reported in the previous version). The signal is within the T90 duration of the Swift BAT GRB, with an estimated "excess" fluence of $F_S(E>0.1 GeV)=1.3 \pm 0.3 \times 10^{-6} erg/cm^2$. This value can be considered the lower limit of the gamma ray fluence in the GeV energy region. However, the Poisson probability of the event to be a background fluctuation is $5.0 \times 10^{-6}$ and it appears in the counting rate of the New-Tupi detector with an annual rate $\sim 76$. In addition, the signal has a significance of only $2\sigma$ in the time profiles with a bin above 2 seconds.  Thus we conclude that the event has a high probability to be background fluctuation. 
\end{abstract}

\pacs{PACS number:  98.70.Rz,  95.85.Pw,  95.55.Ka,  96.50.sf }

\maketitle

\section{Introduction}

The Energetic Gamma Ray Experiment Telescope (EGRET) on board the Compton Gamma Ray Observatory (CGRO) had observed  high energy photon emissions. 
The long duration signals of EGRET in the MeV-GeV range are high energy counterparts of the Burst And Transient Source Experiment (BATSE) GRB signals in the keV-MeV range, being located in the field of view and recorded during the same time periods of the BATSE GRBs \cite{sommer94,dingos94,schneid92,hurley94}. 
These MeV-GeV long duration EGRET emissions arrived at the detector in fragments, and they were delayed or anticipated in relation to the keV-MeV BATSE bursts. 
For instance, in the case of GRB940217, the high energy MeV-GeV emission persisted for at least 5400 seconds, while the low energy sub-MeV emission lasted only 180 seconds.
Nevertheless, it is not clear how high in energy this long duration component could extend, even though many models \cite{totani99,dermer00,pilla98} predict a fluence at GeV-TeV  comparable to that at keV-MeV scales.
It has been also suggested \cite{cheng96} that the fragmentation effect can be a consequence of the electromagnetic cascade, formed by the GeV photons in the infrared-microwave background of the interstellar space. 
There were other gamma-ray bursts (GRBs) detected at high energies, for instance GRB 080514B discovered with the AGILE gamma ray satellite \cite{guiliani08}. 
Since the time of EGRET, this  was the first GRB in which individual photons above several tens MeV have been detected.

On the other hand, the Fermi Large Area Telescope (Fermi　LAT)  is the principal scientific instrument on the Fermi Gamma Ray Space Telescope (Fermi) mission, covering the energy range from about 20 MeV to more than 300 GeV. 
Fermi-LAT has observed gamma ray emissions from the Galactic center \cite{abdo09a}, gamma ray emissions from supernova remnants, blazars and pulsars \cite{abdo09b}, and several GRBs in the MeV-GeV range (Abdo et al. 2009)\cite{abdo09c}. 
In addition, the Fermi Gamma-ray burst Monitor (Fermi GBM) detects $\sim 250$ GRBs (E$\geq$ 50 keV) per year, and about half of them are within the Fermi LAT field of view \cite{ackermann13}. 
However, only $\sim 10\%$ were detected by the Fermi LAT ($\geq$ 100 MeV). 
This means that not all of the GRBs in the keV region have counterparts in the high energy region, above 100 MeV.

Several scenarios have been suggested to explain a possible high energy component, or an extension of the GRB spectrum, such as the synchrotron self-Compton (SSC) model \cite{panaitescu00,kumar08}, that provides a natural explanation for the optical and gamma ray correlation seen in some of GRBs. 
It also implies that a relatively strong second order inverse Compton (IC) component of the GRB spectrum should peak in the energy range of tens GeV \cite{racusin08}. 
Observations by the Tupi experiment can set limits on the strength of this second IC peak.

In this survey we report the occurrence of an excess in the counting rate observed on September 15, 2015. A peak with a duration of $T_{90}= 6.1\pm 0.6)$ seconds and a significance of $4.4\sigma$ could be recognized  by the naked eye in the 24 hours raw data time profile. 
The event triggered the alarm system implemented in the New-Tupi data acquisition scheme. 
In a subsequent analysis we found that the peak onset time was 24.7 s after the Swift BAT GRB 150915A (trigger 655721)  (D'Elia et al. GCN 18315) and within the $T_{90}$ duration of the GRB.  In addition, the BAT trigger coordinates of the GRB150915A has a zenith angle,
$\theta=31.5^0$, relative to the vertical direction of the location of the New-Tupi detector. 

This article is organized as follows: 
Section 2 presents observations showing the association, temporally and spatially, between the New-Tupi counting rate excess and the GRB150915A.
Section 3 is devoted to a confidence analysis, we have examined the time profiles (raw data) through a distribution of the significance level (in standard deviation) for two temporal windows, the full day (September 15, 2015) and for a period of two hours (one hour before and one hour after the GRB150915A trigger time).
In addition, the Poisson probability of the excess being due to a fluctuation of the background is calculated, together with the expected annual rate.
The section, Spectral analysis (presented in the previous version) was eliminated, because we conclude that the event has a high probability to be background fluctuation.

Section 5, we present our conclusions. The paper has a appendix,
devoted to a brief description of the New-Tupi detector, including the data acquisition methodology. 

\section{Observations}

Swift is a first-of-its-kind multi-wavelength observatory designed to the study of GRBs \cite{gehrels04}. 
The Burst Alert Telescope (BAT) on board Swift covers the 15–150 keV energy band and can detect more than 100 GRBs per year. 
The other two instruments are the X-ray telescope (XRT) and the ultraviolet and optical telescope (UVOT).

According to  D'Elia  et al. (GCN 18315), on September 15, 2015 at 21:18:24 UT the Swift BAT triggered and located GRB 150915A (trigger 655721) 
at (R.A., Dec) = (21h 18m 55s, -34d 51' 12''). 
The BAT light curve shows a  long and weak complex structure  with a duration of  $T_{90}=164.7\pm 49.7$ s, in the 15-350 keV energy band and a fluence in the 15-150 keV band of $8.0 \pm 1.8 \times 10^{-7}\; erg/cm^2$ (Palmer et al. GCN 18328). 
In addition, several afterglows linked to GRB 150915A, have been detected in different wavelengths, such as the X-ray afterglow by the Swift-XRT (D'ai  et al. GCN 18329).
However, the Swift-UVOT  instrument (Breeveld  et al. GCN 18322) reported only preliminary $3\sigma$ upper limit.
Optical afterglows were reported by the MASTER II  robotic telescope in South Africa (Rebolo et al. GCN 18316) and an imaging instrument GROND in Chile (Yates et al. GCN 18317).
The optical counterpart of GRB150915A was observed by the European Southern Observatory (ESO) Very Large Telescope (VLT) equipped with the X-shooter spectrograph (D'Elia et al. GCN 18318), and a number of absorption features at a common redshift of $z = 1.968$ were detected and a similar redshift has been found for the GRB/host galaxy redshift.

On the other hand, in most cases the search for GRBs at ground level is to look for an excess of events above the expected background, temporally and positionally coincident with the satellite-detected GRBs, with the aim to obtain the GeV-TeV counterpart of GRBs observed by satellites in the keV-MeV energy band.
However, besides this untriggered analysis, it is possible to make a triggered analysis at ground, similar to the one developed by the Milagro collaboration \cite{milagro}. 
In this method, if an excess above the background is observed, the Poisson probability of this excess being due to a fluctuation of the background is calculated.
Along this line of analysis, a semi-automatic system has been implemented in the New-Tupi detector. 
Every 24 hours, when a day is completed, the data passes through a filter. 
Every time when there is an excess with a significance equal or above 4 sigma in the raw data set, the system triggers an alert. The next task is to verify if the alert is associated with a GRB triggered by satellites.
This is to check whether the GRB coordinates are located  in the field of view of the detector and whether the alert time is within the GRB $T_{90}$ duration.
We have at least 2 alerts per day, and in most cases the alert signals have a duration not longer than 2 s. We have found that these alerts do not have apparent connection (spatially and temporally)
with astrophysics events, and we can consider them with a high confidence background fluctuations.
In fact, we show in Section 3 that they are in the tail of the Gaussian distribution of the background fluctuations.

We believed that the exception was an event  found on September 15, 2015.
The New-Tupi alert system detected  a counting excess, in the scaler mode, with a statistical significance of $6.9\sigma$.
However, after a review of the calculation, the significance of the excess was only $4.4\sigma$. 
The GRB 150915A trigger coordinates were located not far from the vertical direction (zenith) of the New-Tupi location. The situation is summarized in Fig. 1, and represent the equatorial position of the three New-Tupi telescopes, we can see that the  location of the trigger coordinates of the GRB150915A has a zenith angle, $\theta=31.5^0$, relative to the vertical direction of the location of the New-Tupi detector.

\begin{figure}
\vspace*{-0.5cm}
\hspace*{0.0cm}
\centering
\includegraphics[width=10.0cm]{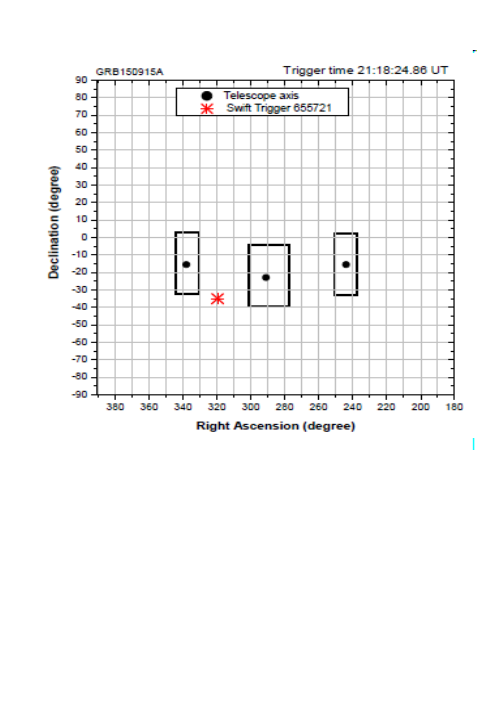}
\vspace*{-5.0cm}
\caption
{
Equatorial coordinate system of a portion of the sky including the field of view of the 
New-Tupi telescopes, represented by rectangles. The asterisk represent  the location of the trigger coordinates of the GRB150915A.   
}
\label{fig1}
\end{figure}

The onset of the excess was at $\sim24$ seconds after the GRB150915A trigger time.
The time profiles of the muon counting rate can be seen in Fig. 2. As already was comment, the signal in the raw data has a significance level of $4.4\sigma$. However, the confidence level of the signal is only around $2\sigma$ in the 4 s, 6 s and 10 s binning counting rates. We highlight, that in all cases, the fluctuation of the background count rate is basically restricted to the $\pm 1\sigma$ region.
\begin{figure}
\vspace*{0.0cm}
\hspace*{0.0cm}
\centering
\includegraphics[width=10.0cm]{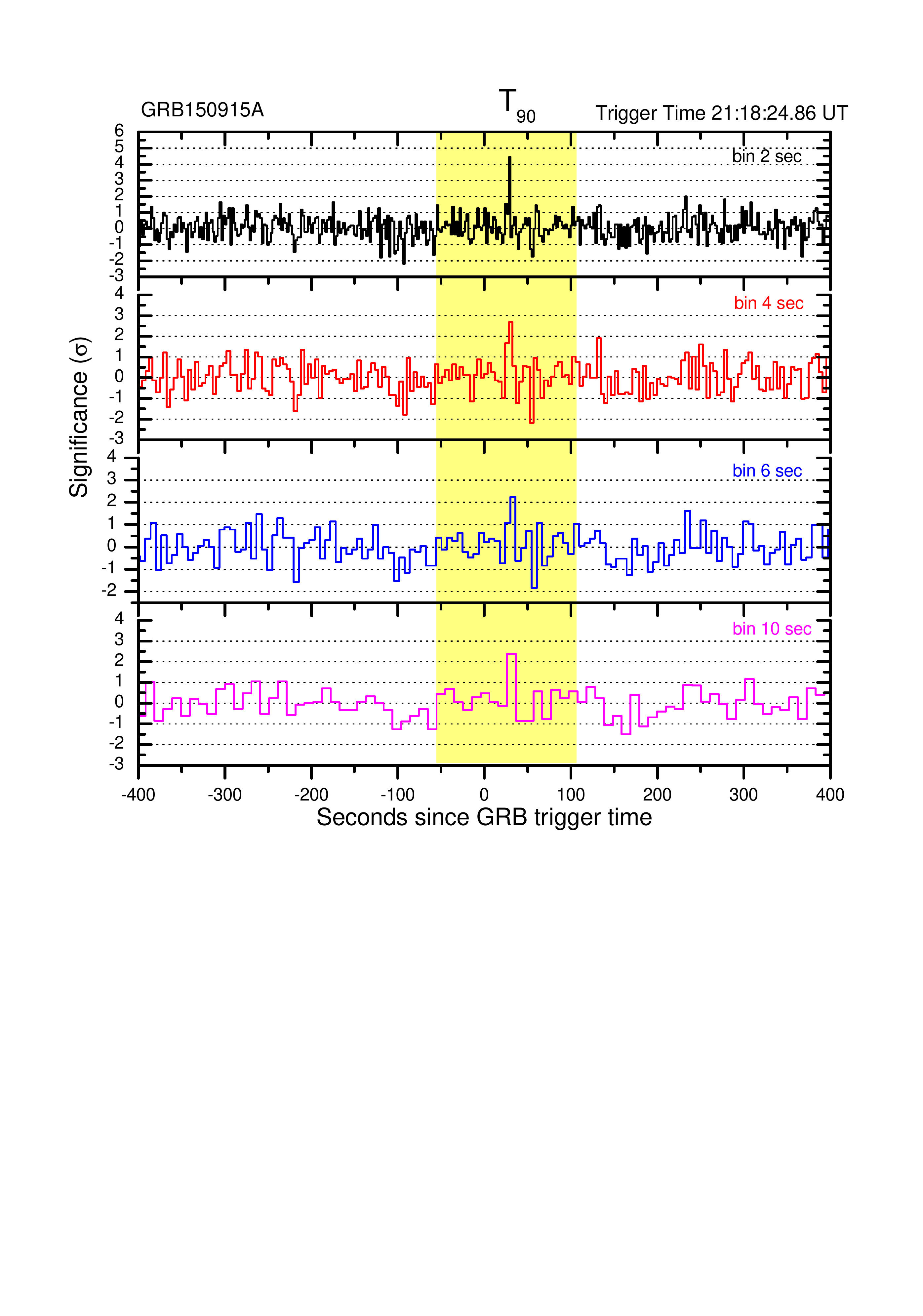}
\vspace*{-5.0cm}
\caption
{
Statistical significance (number of standard deviations) of the 2, 4, 6 and 10 s binning counting rates observed by the New-Tupi detector (scaler mode), as a function of the time elapsed since the Swift BAT trigger time. 
The yellow area corresponds to the T90 duration of GRB150915A.
}
\label{fig2}
\end{figure}

\begin{figure}
\vspace*{0.0cm}
\hspace*{0.0cm}
\centering
\includegraphics[width=10.0cm]{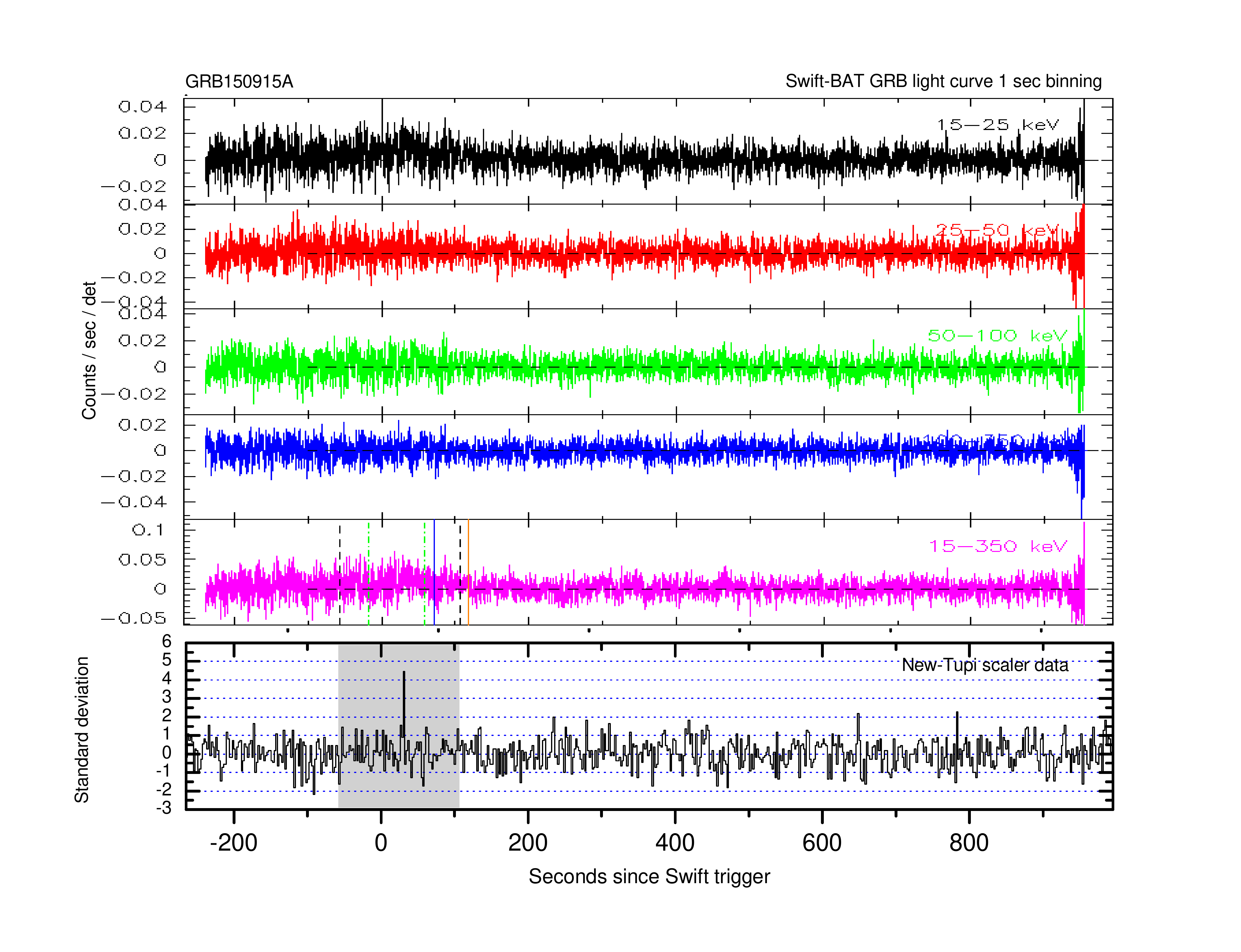}
\vspace*{-0.0cm}
\caption
{
Top five panels: the counting rate of gamma rays in five energy ranges for GRB 150915A observed by the Swift BAT instrument. 
Bottom panel: statistical significance (number of standard deviations) of the 2 s binning counting rate registered by the New-Tupi detector (scaler mode), as a function of the time elapsed since the SwiftGRB 150915A trigger time. 
The gray area corresponds to the T90 duration of GRB 150915A.
}
\label{fig3}
\end{figure}

Fig. 3 and Fig. 4 show a comparison between the time profiles of the Swift GRB140512A event in five energy bands and the statistical significance (number of standard deviations) of the counting rate (2 s and 10 s binning) observed by the New-Tupi detector, as a function of the time elapsed since the Swift GRB140512A trigger time.

\begin{figure}
\vspace*{0.0cm}
\hspace*{0.0cm}
\centering
\includegraphics[width=10.0cm]{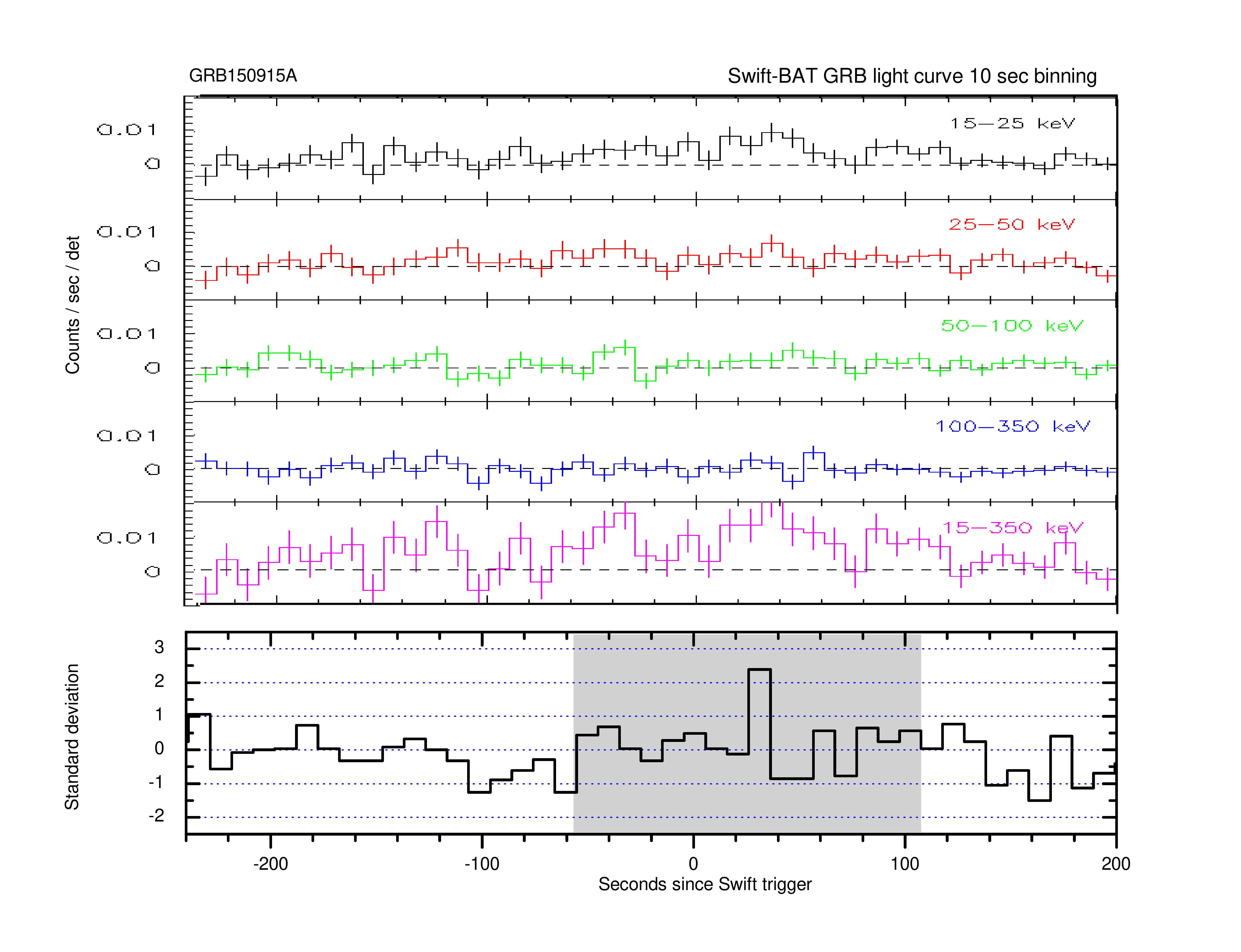}
\vspace*{-0.0cm}
\caption
{
Top five panels: the counting rate of gamma rays in five energy ranges for the event GRB 150915A observed by the Swift BAT instrument. 
Bottom panel: statistical significance (number of standard deviations) of the 10 s binning counting rate registered by the New-Tupi detector (scaler mode), as a function of the time elapsed since the Swift GRB 150915A trigger time. 
The gray area corresponds to the T90 duration of GRB 150915A.
}
\label{fig4}
\end{figure}

\section{Confidence analysis}

\subsection{The ground level data}

In most cases, the counting rate observed at ground level detectors is modulated by the sun activity and by the meteorological conditions, such as the barometric effect. 

 The atmospheric pressure variations are anti-correlated with the flux of secondary particles at ground level, that is,  an increase in the atmospheric pressure is correlated with a reduction in the background rate. The effect arise from the variation of mass of atmosphere above the detector with a periodicity of 12 hours, with peaks at 10am and 10pm in the tropics.

However, as already was commented, at sea level most of particles are muons. The  muon component has an associated barometric coefficient of  up to ten times smaller than the barometric coefficient of NMs and other detectors, most located at high mountain altitudes \cite{kovylyaeva13}, detecting a variety of other particles such as neutrons, protons, electrons, etc. At New-Tupi location the barometric coefficient is about -0.1\% per mb, and
the average the daily barometric pressure variation, peak to peak, that is, from the maximum to the minimum is of up to 3 mb. This means an average daily muon flux variation of up to 0.3\%. Thus the correction on the counting rate of muons by barometric pressure  is not relevant, especially in the analysis of a signal only several seconds to minutes long, very short relative to the periodicity of the pressure variation that is of 12 hours. However, in other ground level detectors such as the NMs this variation can be expressive, about 2\%-3\% in average.  Even so, in a log term variation analysis, it must be included, in all cases,  due to  the tidal effect on atmosphere due to the lunar phases (Full Moon) that introduces an additional modulation with a periodicity 27 days due to the Moon rotation around Earth..

2.- The day/night asymmetry: The cosmic ray in the GeV to $\sim 70$ TeV energy range are also subject to several solar modulations \cite{kane63}, the main are: the day/night asymmetry, with a annual (or seasonal) modulation caused by the annual variation of the rotation axis of the Earth relative to the sun. The daily variation depends of the latitude of location of detector and the effect decreases as the particle energy increases. Thus, for low energy particles (MeV to GeV region), the variations are bigger comparing higher energy particle (TeV region) variations.  

In order to obtain information of this effect on the background counting rate's flux at New-Tupi detector, and to see the stability of the detector, during some days around September 15, 2015, the vertical ``muon'' flux was obtained through the relation, $Flux=Rate/G$, where $Rate$ is the counting rate of the vertical telescope and $G$ is the geometric factor of the vertical telescope, $G=6700 cm^2 sr$ \cite{sullivan71}. The determination of the vertical muon flux, will be useful also in the spectral analysis, and that will be addressed in the next section.

The result is shown in Fig. 5 for the available data of September of 2015, taking into account the count efficiency of 
95\% of the telescope. We have included a comparison with the time profiles of the count rate (pressure corrected) of the Thule Neutron Monitor, located at high latitude (76.5N, 68.7W) and 26m asl. Despite the high differences between New-Tupi and Thule latitudes, the Thule's longitude is only 1.6h delayed relative to the New-Tupi's longitude. This means that the phase of the daily variation between both detectors must be small. Indeed, we can see a good correlation between both, as shows in Fig. 5. We would like to point out, that the data period in Fig. 5 is close to the equinox, where the Earth's equator is parallel to the ecliptic plane and the duration of day and night is the same
in both hemispheres at equivalent latitude. 

\begin{figure}
\vspace*{-0.0cm}
\hspace*{0.0cm}
\centering
\includegraphics[width=13.0cm]{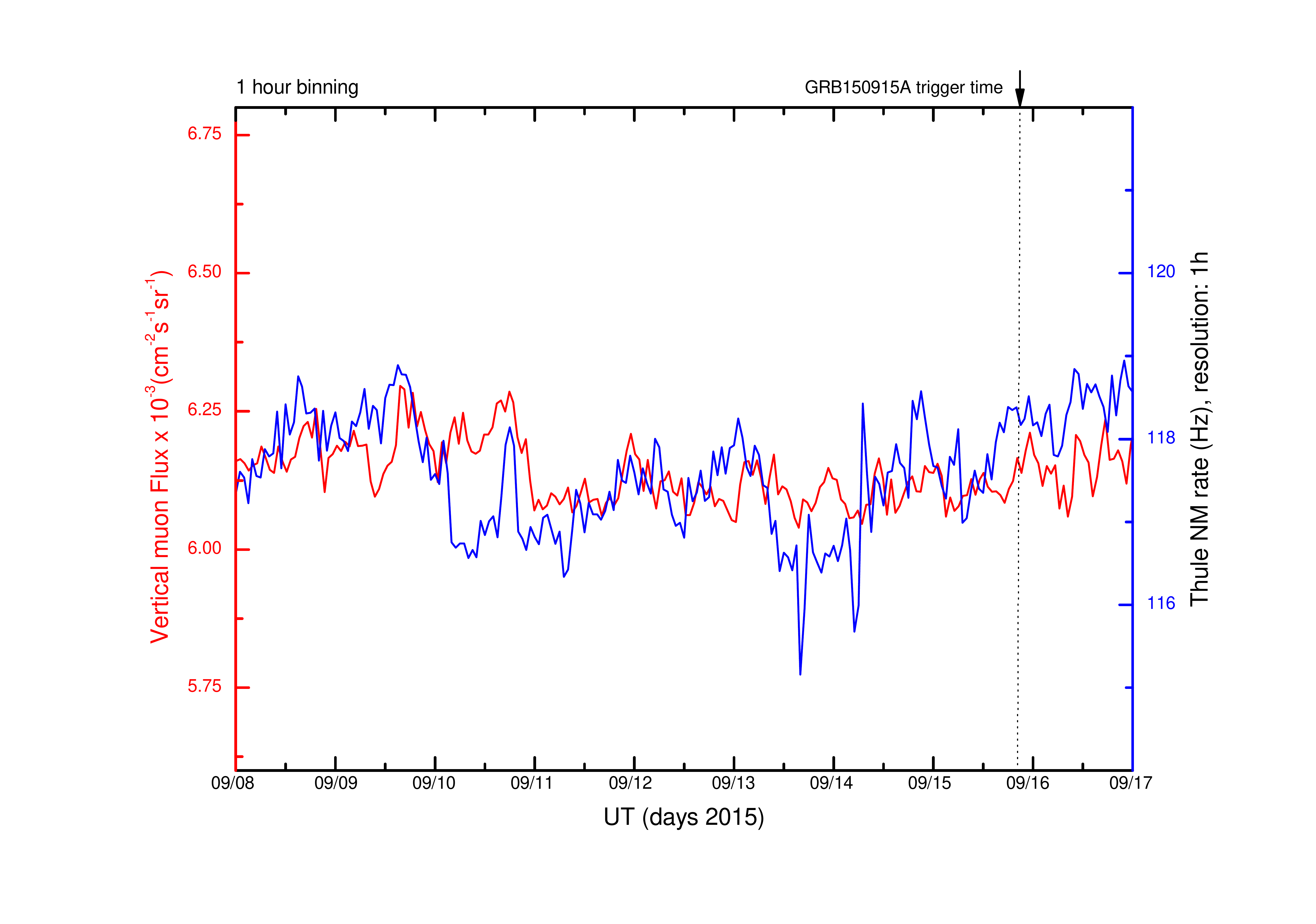}
\vspace*{-1.0cm}
\caption
{
Left scale: linked with the red curve, represents the vertical muon flux as measured by the vertical New-Tupi telescope, for nine consecutive days with beginning on September 8, 2015 at 00:00 UT 
Right scale: linked with the blue curve, represents the counting rate as measures by the Thule NM and for the same period. The vertical line on September 15 indicates the occurrence of the GRB150915A trigger time.
}
\label{fig5}
\end{figure}

Following Fig. 5, we can see that in most cases,
at New-Tupi location the daily variation (peaks) has a maximum at about $\sim 18h$ UT (15h local time), with variations of up to 1.5\%. However, the average is less than 1\% and present daily variations.

There are also several long term modulations, for instance, the solar wind velocity shown a strong  27-day recurrence tendency  due to solar rotation \cite{alania10}  and that has influence in the geomagnetic field and consequently in the low energy cosmic rays. There are also very long term modulations, such as the annual variation and the main is the variation due to solar cycle effect \cite{mccracken03}, that modulates the flux with a periodicity of 11 years. These variations are irrelevant in the analysis of a signal only several seconds long, relative to a background rate around of signal.


The New-Tupi detector has operates with a duty cycle of 76\% and it has registered three of the four stronger geomagnetic storms of the current solar cycle, as well as, several other the minor intensity, the effect at ground level is a fall in the counting rate known as Forbush decrease. The New-Tupi observations correlated with the geomagnetic indices such as the planetary Kp index that characterize the magnitude of geomagnetic storms. The Kp is in the range 0–9 with 1 to 4 being calm and 5 or more indicating a geomagnetic storm. The New-Tupi observations are also correlated with other observations made by other ground level detectors, such as the South Pole Neutron Monitor. Fig. 6 shows the ground level effect of the second higher geomagnetic storm in the current solar cycle, known as the ``solstice storm''. This geomagnetic storm reached the condition of $Kp=8$ or ($G3=8$ severe) in the NOAA storm scale.  The analysis of this and the other geomagnetic storms are in progress.

\begin{figure}
\vspace*{-0.0cm}
\hspace*{0.0cm}
\centering
\includegraphics[width=8.0cm]{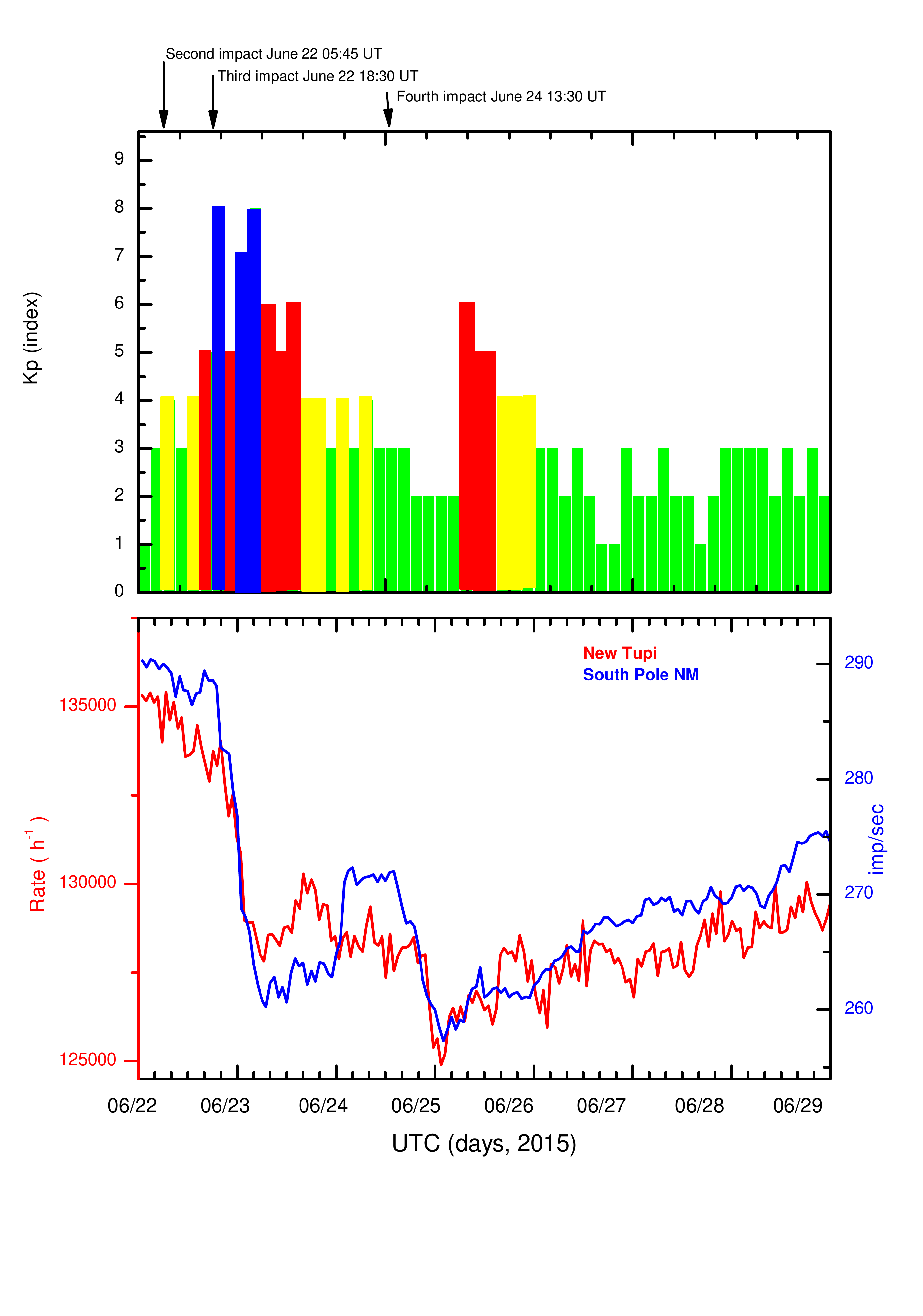}
\vspace*{-1.0cm}
\caption
{Top panel: Estimated planetary $K_P$ index (3 hour data) for seven consecutive days with beginning
on June 22, 2015 at 00:00 UT. Bottom panel:
Time profiles of the counting rate, for the same period, observed in two different detectors, the red curve correspond to the New-Tupi vertical telescope, and the blue curve correspond to the South Pole Neutron Monitor. The vertical arrows indicate  the arrival time on Earth of three of four  coronal mass ejections (CMEs)  in the period, triggering geomagnetic storms.
}
\label{fig6}
\end{figure} 

The Tupi detector has registered also several particle excesses of upto 4\% relative to the muon background during the occurrence of ``radiation storms'', that is, relativistic particles (protons) accelerated by coronal mass ejection (CMEs), and producing secondart particles in the Earth atmosphere. The typical duration of these excesses is of several hours. The observations are in association with the increase of the proton flux at GOES satellite, as well as with other ground level observation, an example is reported in \cite{augus16}.

\subsection{New-Tupi data on September 15, 2015}

September 15, 2015 was a typical almost quiet day, only three solar sunspots visible on the solar disk, and of them, only the region AR2515, at East of solar disc, 
had a small activity, emitting C-class (minor) flares in the morning.

In order to see with more accuracy the background rate
fluctuation, we have examined the time profiles (raw data) through a distribution of the significance level (in standard deviation) for the full day (September 15, 2015) and that is shown in Fig. 7 (top panel). 

\begin{figure}
\vspace*{-0.0cm}
\hspace*{0.0cm}
\centering
\includegraphics[width=8.0cm]{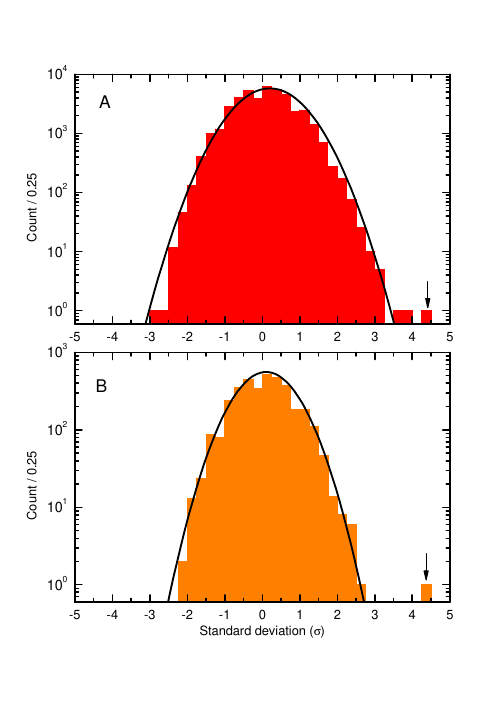}
\vspace*{-1.0cm}
\caption
{
Distribution of the fluctuation counting rate (raw data) for the New-Tupi
detector in scaler mode (in units of standard deviations) for two temporal windows.
Top panel: for the full day period September 15, 2015. Bottom panel: inside 2 hr
interval (1 hr before and 1 hr  after the trigger) around the Swift
GRB150915A trigger time. The signal with a significance of $4.4\sigma$ (marked by vertical arrows)represents the New-Tupi signal associated with the Swift GRB. In both cases, the black curves, represent a Gaussian fit.
}
\label{fig7}
\end{figure}

In Fig. 7, the arrows (out of the Gaussian distribution indicated by a bold black line) represent the signal (peak) that correlates with the GRB150915A. However, there is at least
a peak with a confidence of up to $4\sigma$. However,  this event is very close to the tail of Gaussian distribution of the background rate, this means that the probability to be a background fluctuations is high.

The situation is better when the background rate distribution is obtained only for one hour before and one hour after the trigger time of  GRB150915A, as is shown 
in Fig. 7 (bottom panel). In this case the signal is more evident and far from background rate distribution.

On the other hand, if the muon excess observed by the New-Tupi detector is due to the background fluctuations, then the following conditions are met:
a) The number of excessive counts in two disjoint time intervals is independent.
b) The probability of an excess during a small time interval is proportional to the entire length of the time interval.
Under these conditions, the Poisson distribution allows to obtain the frequency of excess events at a chosen significance level in a given time interval.
This method is useful to determine the expected annual rate of a given muon excess  to be a background fluctuation.

Fig. 8 shows the Poisson distribution to the counting rate on September 15, 2015 (red circles). As the count was read at every 2.038 s, the distribution
was mounted with $N=42400$ points and binning with a width of $\Delta \log P=0.2$.
 We also include for comparison the Poisson distribution to eleven quiet days (black curves), without  storms. These black curves were generated by connecting points with lines, sequentially.

The excess, n, observed in the counting rate, associated to the GRB150915A (vertical arrow) be background fluctuation is $\log_{10}(P(n))=-5.3$, or, $P(n)= 5 \times 10^{-6}$,  and the frequency with that appear in one day, can be obtained as
\begin{equation}
f(n)=NP(n)=0.21.
\end{equation}
This means an  annual rate of $\sim 76$

\begin{figure}
\vspace*{0.0cm}
\hspace*{0.0cm}
\centering
\includegraphics[width=14.0cm]{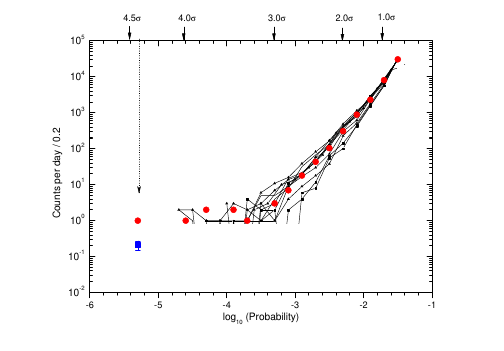}
\vspace*{-0.5cm}
\caption
{
The Poisson probability distribution of a signal to be a background fluctuation 
in the counting rate (raw data) observed by the New-Tupi detector. The black curves represent the data of 11 typical calm days (without storms). These black curves were generated by connecting points with lines, sequentially.
The red circles: represent the data on September 15, 2015. The Blue square indicates the number of times that the signal associated to the GRB150915A (vertical dotted arrow) appears per day, in the case to be background fluctuation. The statistical significance is indicated by the arrows at top.
}
\label{fig8}
\end{figure}

The results as shown in Fig. 7 (top panel) and Fig. 8, means that events with significance around $4\sigma$
are common in the raw data. This behaviour already has been observed in the analysis of two events in our 
previous paper \cite{augusto15}. One way to avoid the fluctuations on the raw data is to make the analysis in the time profiles made with a temporal bin with width larger (like 5 sec binning).  Because the significance even for bin 5 seconds, continued above $4\sigma$. However, in the present case, the significance is only around 2 sigma when is used a bin of 5 seconds.
Thus, we conclude that the probability of the excess (in apparent association with the GRB150915A), be background fluctuation is high.

\section{Conclusions}

We carried out a systematic search for a GeV counterpart observed at ground level
of GRBs triggered in gamma ray detectors on-board of satellite. The search is made
since 2014, using a new detector (New-Tupi) consisting of four plastic scintillators ($150 \times 75 \times 5 cm^3$) placed in pairs and located in Niteroi, Rio de Janeiro, Brazil.

On September 15, 2015 at 21:18:58 UT,
the New-Tupi alert system, triggered an increase in the counting rate (excess), with 
the following characteristics: \\
a) A statistical significance of $4.4\sigma$ (scaler mode).\\ 
b) A duration of  $T_{90}=6.1$ s. \\
c) The excess has an onset at $T_0+24.7$ s, where $T_0$ is 
the the GRB150915A trigger time, this means that the excess correlates temporally with the GRB150915A, because it is within the T90 duration of the Swift GRB. \\
d) The trigger coordinates of the GRB150915A has a zenith angle of $\theta=31.5^0$, relative to the vertical of New-Tupi location. This means that the excess correlates spatially with the GRB150915A, because the source of gamma rays producing muons in the atmosphere is in the field of view of the New-Tupi detector. \\
e)The signal in the raw data,  persists above the background with a confidence level greater than $2\sigma$ in the 4 s, 6 s and 10 s binning counting rates, while the oscillation of the background in most cases is only around $\sim 1\sigma$.\\

The second task was a confidence analysis, with the following results:\\
a) The counting rate in the New-Tupi detector was made with stability with a duty cycle of 
76\%. The counting rate correlates with solar transient events, for instance, we have register with high confidence, three of the four stronger geomagnetic storms of the current solar cycle. The Forbush decrease correlates with the geomagnetic indices and other ground level observations and so on.\\
b)The muon flux derived from the counting rate is in agreement with other observations at equivalent latitude and at sea level. In addition, it is subject to solar modulation, the daily variation or the day/night anisotropy appear in the data, and correlated with observations made, approximately at the same longitude. \\
c)The distribution of the fluctuation counting rates  were derived by two temporal windows, the first consider the full day (September, 15, 2015) and the second consider the data only one hour before and one hour after the GRB150915A trigger time. Both follow the expected Gaussian distribution. In both also the signal associated to the GRB  appear isolated and out of Gaussian distribution. However, In the first case, we can see a very small number of fluctuations in the counting rates, and some of them, with a significance of  up to $4\sigma$, and they are present even on calm days (without storms), they are in the tail of the distribution,  and as they do not have apparent connection with astrophysics events, we can consider them with a high confidence background fluctuations.\\ 
d)The chances of the event to be a background fluctuation is $5.0 \times 10^{-6}$ and it appears in the counting rate of the New-Tupi detector with an annual rate $\sim 76$.\\

In addition, the significance of the signal is only around $2\sigma$ in the time profiles binning with a temporal width above 2 sec. Thus, the event observed in apparent correlation with the GRB150915A has a high probability to be background fluctuation.

\appendix
\section{The New-Tupi detector} 
\label{App:AppendixA}

The aim of the New-Tupi detector is to study space weather effects driven by diverse solar transient events.
In addition, the New Tupi detector is used to search for the GeV counterparts of GRBs observed by spacecraft detectors. 
The New-Tupi detector is located in Niteroi city, Rio de Janeiro state, Brazil ($22^0 53'00''S,\; 43^0 06'13'W$, 3 m above sea level), that is  within the South Atlantic Anomaly (SAA) region,  where the inner Van Allen radiation belt makes its closest approach to the Earth’s surface. 
As a consequence, the SAA region is characterized by an anomalously weak geomagnetic field strength (less than 28,000 nT) \cite{barton97}. 
In this way, the shielding effect of the magnetosphere is not perfectly spherical and has a ``hole'' as a result of the eccentric displacement of the center of the magnetic field.

The New-Tupi detector consists of four units assembled on the basis of a slab (150cm x 75cm x 5cm)  of plastic scintillator   and a Hamamatsu photomultiplier  of 127 millimeters in diameter, with 10 stages.
Each unit is placed inside a box with a truncated square pyramid shape (see Fig. 9). 

The data acquisition system is made on the basis of a  data acquisition (DAQ) device with 8 analog channels, 1 MS/s per channel, 12-bit resolution, with simultaneous sampling rate. 
Only signals above a threshold value, corresponding to energy of about 100MeV deposited by particles (mostly muons) that reach the detector, are registered. 
The output raw data of each detector is recorded at a rate of one Hz.
The GPS receiver (New-Tupi) outputs Universal Time (UT).

\begin{figure}
\vspace*{-4.0cm}
\hspace*{0.0cm}
\centering
\includegraphics[width=13.0cm]{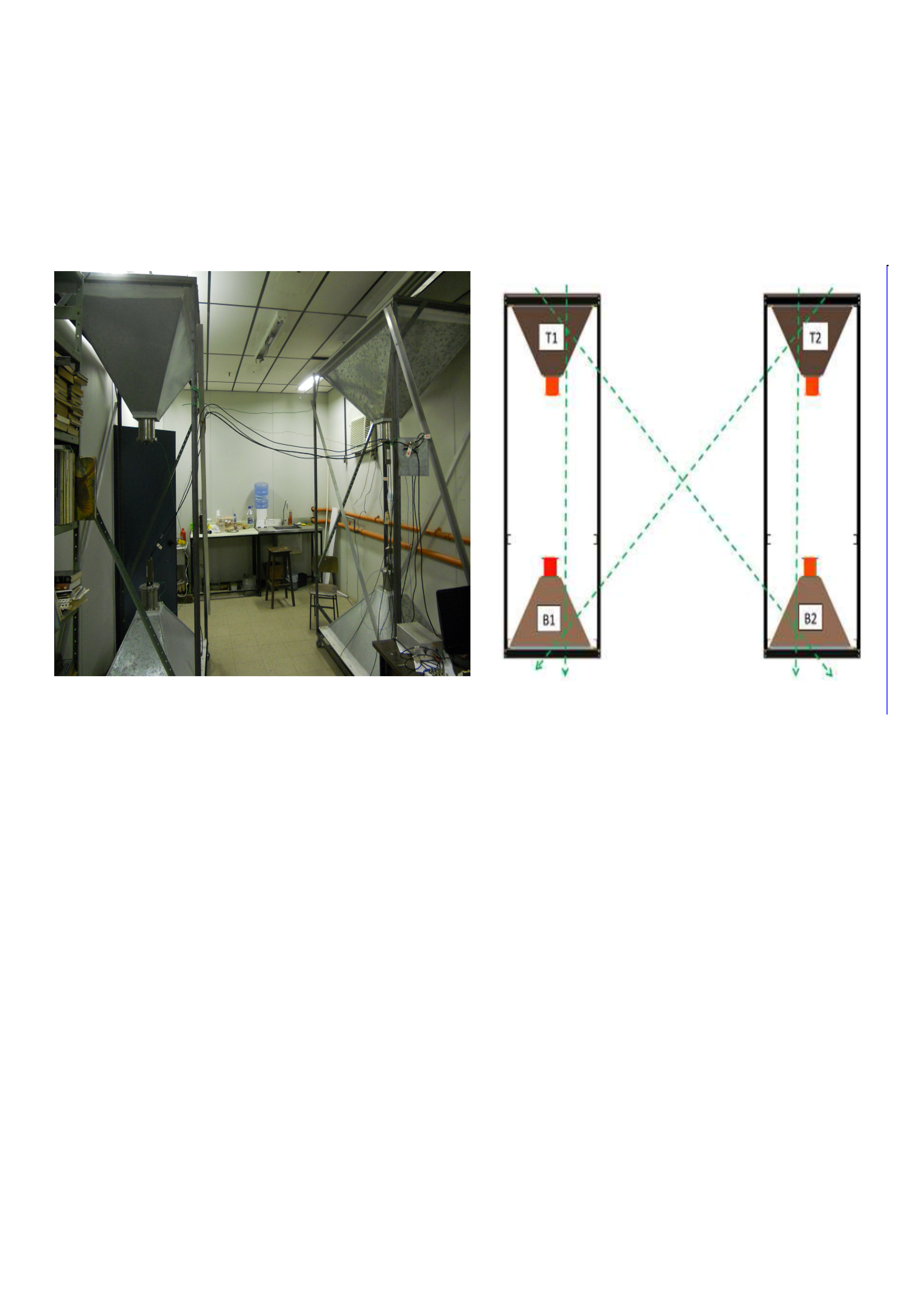}
\vspace*{-9.0cm}
\caption
{
Left: Photograph of the New-Tupi detector.
Right: Scheme of the New-Tupi detector. 
The detector configuration in the telescope mode of operation (counting the coincidence pulses from two PMTs)  allows to measure the muon flux from three directions, the vertical (zenith), west and east (the last two with an inclination of 45 degrees to the vertical, as is indicated by the dashed lines).
The scaler mode consists in recording the single hit rates of all four PMTs at a fixed rate.
}
\label{fig9}
\end{figure} 

\subsection{New-Tupi in the telescope mode}

Fig. 9 (left) shows a photography of the New-Tupi detector. 
The four detector units are placed in pairs, with T1 (top) and B1 (bottom) and T2 and B2 respectively, as shown in Fig. 9 (right).
This layout allows to obtain the muon flux from three directions, the vertical, west and east, the last two with an inclination around 45 degrees. 
The telescopes registers the coincidence rate between T1 and B1; T2 and B2  (vertical), T1 and B2 (west), and T2 and B1 (east).
Both vertical and lateral separation between the units are 2.83 m.

\subsection{New-Tupi in the scaler mode}

In parallel with the telescope mode, New-Tupi is operated in the scaler mode (or single particle technique) \cite{obrian76,morello84,aglietta96}, where the single hit rates of all of the four PMTs are recorded once a second.
This mode allows to search for excess of muon flux at fixed time intervals, even when the source is out of the field of view of the telescopes.
This is because the field of view of a single unit is wider than the field of view of the whole detector.
However, the efficiency of particle detection in the scaler mode decreases as the zenith angle of the incident particle increases, due to atmospheric absorption. 
Thus, the scaler method is limited to incident particles with the zenith angle less than 60 degrees.

\acknowledgments

This work is supported by the National Council for Research (CNPq) of Brazil, under Grant  No. $306605/2009-0$ and Fundacao de Amparo a Pesquisa do Estado do Rio de Janeiro (FAPERJ) under Grant No. $08458.009577/2011-81$ and E-26/101.649/2011.
This research has made use of data provided by the Swift team.
We  express our gratitude to the Goddard Space Flight Center for valuable information and data used in this study through the
GCN web page ($http://gcn.gsfc.nasa.gov/$).

\end{document}